\documentstyle[12pt,epsf]{article} 
\font\tenrm=cmr10 

\newcommand{\bref}[1]{(\ref{#1})} 
\newcommand{\ct}[1]{\cite{#1}}

\newcommand{\be}{\begin{equation}} 
\newcommand{\ee}{\end{equation}} 

\def\theequation{\thesection.\arabic{equation}}
\def\@eqnnum{{\rm (\theequation)}}

\def\lsim{\mathrel{\rlap{\lower4pt\hbox{\hskip1pt$\sim$}}
    \raise1pt\hbox{$<$}}}
\def\gsim{\mathrel{\rlap{\lower4pt\hbox{\hskip1pt$\sim$}}
    \raise1pt\hbox{$>$}}}
\def\frac#1#2{{{#1} \over{#2}}} 

\begin{document}  
\begin{titlepage} 
  
\begin{flushright}  
{CCNY-HEP-96/16 \\ }   
{CU-TP-796 \\ }
{MPI-PhT/96-119 \\ }
{hep-lat/9704016 \\ }  
{\hfill November 1996 \\ }  
\end{flushright}  
\vglue 0.2cm  
	   
\begin{center}   
{ 
{\Large \bf Monte Carlo Studies of \\ 
Two-Dimensional Systems with a $\theta$ Term \\ }  
\vglue 1.0cm  
{Jan C.\ Plef\/ka $^{a\ 1}$ \\ }   
\vglue 0.25cm  
{and} 
\vglue 0.25cm  
{Stuart Samuel $^{bc\ 2}$ \\ }   
\vglue 0.5cm  

{\it $^{a}$Department of Physics\\}
{\it City College of New York\\}
{\it New York, NY 10031, USA\\} 
\vglue 0.6cm  
{\it $^{b}$Department of Physics\\}
{\it Columbia University\\}
{\it New York, NY 10027, USA\\} 
\vglue 0.6cm  
{\it $^{c}$Max-Planck-Institut f\"ur Physik\\}
{\it Werner-Heisenberg-Institut\\} 
{\it F\"ohringer Ring 6\\} 
{\it 80805 Munich, Germany\\}

\vglue 0.8cm

  
{\bf Abstract} 
} 
\end{center}  
{\rightskip=3pc\leftskip=3pc 
\quad A $\theta$ term, 
which couples to 
topological charge, 
is added to the two-dimensional 
lattice $CP^{3}$ model 
and $U(1)$ gauge theory.  
Monte Carlo simulations are performed and 
compared to strong-coupling character expansions.   
In certain instances,  
a flattening behavior occurs 
in the free-energy at sufficiently large $\theta$, 
but the effect is an artifact 
of the simulation methods.  

}

\vfill

\textwidth 6.5truein
\hrule width 5.cm
\vskip 0.3truecm 
{\tenrm{
\noindent 
$^{1)}$ \hspace*{0.2cm}E-mail address: t52@next.nikhef.nl \\ }
$^{2)}$\hspace*{0.35cm}E-mail address: samuel@scisun.sci.ccny.cuny.edu \\ }
 
\eject 
\end{titlepage}

\newpage  

\baselineskip=20pt

{\bf\large\noindent I.\ Introduction}\vglue 0.2cm
\setcounter{section}{1}   
\setcounter{equation}{0}   

Formally, 
the inclusion of a $\theta$ term 
in a theory 
does not affect 
the equations of motion.  
For this reason, 
it was once thought 
that $\theta$ terms were irrelevant. 
In fact, 
this appears to be the case 
for the four-dimension $U(1)$ gauge theory.  
Adding the $\theta$ term 
$\int d^4 x F_{\mu \nu} \tilde F^{\mu \nu} (x)$ 
to the action 
is believed to not affect physics.  
The situation for a non-abelian theory 
is  different.  
The importance of the $\theta$ term 
$ 
  S_{\theta} = 
   g^2 \theta \int d^{4} x  F_{\mu \nu}^a 
      \tilde F^{\mu \nu}_a (x) / (32 \pi^2) 
$ 
was realized when instantons 
were uncovered in 
four-dimensional Yang-Mills theories 
\ct{bpst75}.  
Instantons 
represent barrier-penetration processes 
between different classical $n$-vacua.   
An $n$ vacuum $| n \rangle $ 
is obtained from the perturbative vacuum 
through a non-trivial gauge transformation 
carrying winding number $n$.  
Instanton-barrier-penetration effects 
imply that the true vacua are linear combinations 
of $n$ vacua.  
These vacua $| \theta \rangle$ 
are called $\theta$-vacua 
\ct{cdg76,jr76}  
and are given by 
$
  | \theta > = \sum_{n=-\infty}^{\infty} 
  \exp \left( { i n \theta } \right) | n \rangle 
$.  
In a functional integral, 
$\theta$-vacua are incorporated by adding 
the term $S_{\theta}$ to the action.  
Since $S_{\theta}$
breaks parity, 
time-reversal invariance 
and CP symmetry  
when $\theta \ne 0$ or $\theta \ne \pi$, 
the strong interactions 
explicitly violate these symmetries 
for $0 < \theta < \pi$.  
In fact, 
not only do instantons contribute to $S_{\theta}$
but all topological quantum fluctuations 
also do so. 
Such topological fluctuations are known 
to exist because they contribute significantly 
to the $\eta^\prime$ mass 
\ct{fgl73}--\ct{cdpv90}.  

In QCD, 
phases in the quark mass matrix ${\cal M}$ 
also contribute to CP violation.  
However, 
only the combination 
$\theta_{eff} = \theta + Arg Det {\cal M}$ 
is relevant for strong CP violation 
for the following reason.   
Through redefinitions of quark fields, 
one can eliminate 
the CP violating phases in the quark mass matrix.  
However, 
to eliminate one CP-violating phase, 
a $U(1)$ axial rotation is used.  
Via the axial anomaly, 
$Arg Det {\cal M}$ re-emerges 
as a coefficient of $ S_{\theta}$.  
Hence, 
the physical effective theta parameter 
is $\theta_{eff}$.   
The strongest constraint on $\theta_{eff}$ 
comes from the electric dipole moment of the neutron.   
Compatibility with experimental bounds requires
$\theta_{eff} \lsim 10^{-9}$ 
\ct{baluni79,cdvw79}.  
The strong CP problem in QCD can then be phrased as 
the question of how $\theta_{eff}$ 
can naturally be so small.  
The strong CP problem is thus a fine-tuning issue. 

In a pure Yang-Mills theory, 
the strong CP problem  
involves vacuum dynamics.  
Vacuum physics is related to the long-distance behavior 
of a theory 
and hence non-perturbative and strong-coupling effects.  
For four-dimensional Yang-Mills theories, 
this regime 
is not well understood.  
Thus
it is useful to consider simpler systems, 
such as 
the two-dimensional $CP^{N-1}$ models.  
They have features in common with 
four-dimensional Yang-Mills theories: 
They are asymptotically free, 
possess instanton solutions, 
and have $\theta$ vacua.  

Insight into the $CP^{N-1}$ models 
comes from strong coupling expansions 
\ct{rs81}-\ct{ps96a}, 
Monte Carlo simulations 
\ct{hm92}-\ct{schierholz94}, 
and the use of the large $N$ limit \ct{dlv78,witten79b}.  
As $N \to \infty$, 
the $CP^{N-1}$ model becomes 
a system of $N$ free particles 
and $N$ free antiparticles,   
and there is no $\theta$ dependence: 
$\theta$-vacua are degenerate in energy.   
However, 
the first $1/N$ correction 
lifts the degeneracy of the $\theta$-vacua, 
with the energy separation between $\theta$-vacua 
being of order $1/N$. 
The first $1/N$ correction also leads to 
a quantum-mechanically generated $U(1)$ gauge field.  
This gauge field produces 
a linear potential between  
particles and antiparticles.  
Hence the system confines, 
although the string tension, 
in this approximation, 
is again of order $1/N$.   
Actually, 
it is expected that the confining force becomes stronger 
and stronger as higher-order $1/N$ corrections are included 
since the $CP^{N-1}$ models exhibit superconfinement 
\ct{samuel83}. 
The dramatic changes in behavior 
in going from leading order to next-to-leading order 
are indicative of the singular nature of the $1/N$ expansion 
in the $CP^{N-1}$ models.  
The singular nature is also reflected 
in the fact that strong coupling and $1/N$ expansions 
do not commute 
\ct{rs81,samuel83,hhr80}.    
For this reason, results from Monte Carlo methods should 
probably be trusted over results from $1/N$ methods. 

In the presence of a $\theta$ term, 
G. Schierholz and co-workers in pioneering work  
have performed Monte Carlo simulations 
on the  $CP^{3}$ model.\ct{schierholz94}  
Results for the free energy 
were interesting and unexpected.  
For fixed inverse coupling $\beta$, 
a dramatic change in the free energy behavior 
occurred at a critical value $\theta_c$ of $\theta$.
The free energy $f$ per unit volume 
was well represented by 
\be
  f =
    \left\{ \begin{array}{ll} 
    a\left( \beta  \right) \theta^2 
       \quad &\theta < \theta_c \\ 
    c\left( \beta  \right) \quad \, \quad &\theta > \theta_c   
            \end{array} 
     \right. 
\quad . 
\label{eq1} 
\ee
Hence, 
for $\theta \ge \theta_c$, 
the free energy had no dependence on $\theta$.  
In two dimensions, 
the string tension $\sigma (e,\theta)$ 
for external particles of charge $e$ 
in a theta vacuum $| \theta \rangle$ 
can be computed from the free energy using 
$ 
  \sigma (e,\theta) = 
  f(\theta + 2\pi e) - f(\theta) 
$.\ct{luscher78}   
Hence, 
for $\theta > \theta_c$,
the string tension vanishes for 
particles of sufficiently small charge.   
In other words, 
confinement for small external charges is lost.  
Furthermore, 
the simulations 
in \ct{schierholz94} 
indicated that $\theta_c$ went to zero 
in the continuum limit in which  
the coupling $g$ goes to zero. 

In \ct{ps96a}, 
strong coupling character expansions 
of the free energy were obtained.  
Near infinite coupling, 
no flattening behavior of the free energy 
like that 
of Eq.\ \bref{eq1} 
was seen. 
However, 
at smaller $g$, 
a peak in the free energy occurred.  
Although higher corrections might change this behavior, 
the strong coupling series 
provided some support 
for the form of the free energy given 
in Eq.\ \bref{eq1}.  

Many two-dimensional systems 
with a $U(1)$ gauge field,  
such as the Schwinger model, 
possess a cusp in the free energy at $\theta = \pi$.  
The cusp signals the spontaneous breaking 
of CP invariance.\ct{coleman76}  
At infinite coupling, 
the lattice $CP^{N-1}$ models 
also undergo spontaneous CP breaking 
at $\theta = \pi$.\ct{seiberg84}   
Hence, a phase transition 
in the $CP^{N-1}$ models at $\theta_c = \pi$ 
is not unexpected.  
What is interesting about the numerical studies 
\ct{schierholz94} 
of the $CP^{3}$ model 
is that, for sufficiently weak coupling, 
$\theta_c$ moves away from $\pi$ 
and decreases with decreasing $g$.  
This picture suggests that, 
to obtain a continuum confining theory  
from the lattice $CP^{3}$ model, 
one must tune $\theta$ to zero.  
This then also suggests 
that continuum {\it confining} $CP^{N-1}$ models 
must have $\theta = 0$.%
{\footnote{One would still need to explain 
why confinement is necessary 
for the continuum limit.}}   
If the analog of this statement were true 
for a four-dimensional Yang-Mills theory, 
then the strong CP problem would be solved.  
In fact, 
preliminary studies 
\ct{schierholz95} 
of the free energy 
of the four-dimensional $SU(2)$ Yang-Mills theory  
show a free energy behavior similar 
to the one in equation 
\bref{eq1}.  
However, 
in four-dimensional Yang-Mills theories, 
the string tension is not related to the free energy $f$.  
Hence, 
the analog of the argument for the $CP^{3}$ model, 
namely, 
that $\theta$ must be tuned to zero to obtain confinement, 
does not necessarily hold.  
Nonetheless, if $\theta$ must be less than $\theta_c$ 
for some other physical reason, 
and if $\theta_c$ goes to zero as $g \to 0$, 
then the strong CP problem would be solved 
in the Yang-Mills theories.  
This is interesting 
because 
it has been suggested 
that the strong CP problem might be solved 
naturally within the pure 
Yang-Mills theory.\ct{samuel92} 
In fact, 
such a result occurs 
in a $2+1$ dimensional model: 
the Yang-Mills sector generates 
a relaxation field, 
which acts like the axion 
in the Peccei-Quinn mechanism 
\ct{pq77,weinberg78,wilczek78}.  
In ref.\ct{samuel92}, 
criteria were established 
to determine when 
such a natural relaxation mechanism arises 
for theories in arbitrary dimensions.  
It has not been yet possible to determine whether  
these criteria are satisfied 
for four-dimensional Yang-Mills theories.  

One unusual feature 
of the Monte Carlo simulations of 
ref.\ \ct{schierholz94} 
is the volume dependence 
on the free energy 
for sufficiently large $\theta$. 
As the volume $V$ increases, 
$\theta_c$ decreases.  
This leads to a situation in which 
the free energy differs by large factors 
as $V$ varies.  
For example, 
at $\beta = 2.7$
and $\theta$ near $\pi$, 
the free energy on a $46 \times 46$ lattice 
was three times 
the free energy on a $72 \times 72$ lattice.   
See Figure 3 
of ref.\ \ct{schierholz94}. 
Since quantities, 
like the free energy, 
should quickly converge 
in the thermodynamic limit, 
there are only three possible explanations:  

\begin{itemize}
\item[(i)] For sufficiently large $\theta$, 
there are light, perhaps massless, 
modes in the system 
which cause finite-size effects.  
\item[(ii)] A systematic effect occurs 
which leads to numerical results 
that differ from true results. 
\item[(iii)] There is some unknown explanation 
not covered by (i) or (ii).  
\end{itemize} 

One purpose of the current work 
is to try to determine whether (i) or (ii) occurs.  
We perform Monte Carlo simulations 
on an exactly solvable $U(1)$ lattice gauge theory.  
By comparing numerical and analytic results, 
much insight into simulating systems 
with a $\theta$ term is gained.  
Our results are presented in Section III. 
We will argue that systematic effects 
can lead to anomalous flattening behavior 
in the free energy, 
as described by 
Eq.\ \bref{eq1} 
for the $U(1)$ lattice gauge theory.   
In Section II, 
a general analysis 
of simulating systems with a $\theta$ term 
is presented.  
Section II 
provides a mechanism 
by which anomalous flattening behavior can arise. 
In Section IV, 
a lattice $CP^3$ model is treated.  
By comparing Monte Carlo data 
with analytic strong coupling series 
from ref.\ \ct{ps96a}, 
flattening behavior can be shown to be anomalous 
for at least two simulations.  
Some additional results and 
remarks are presented in Section V.  
In Section VI, 
a summary and final discussion is given.

\bigskip 
{\bf\large \noindent II.\ General Issues Concerning Simulations 
with $\theta$ Terms}  
\setcounter{section}{2}   
\setcounter{equation}{0}   

In this section, 
we discuss some issues concerning simulations 
of an arbitrary lattice theory 
in the presence of a $\theta$ term.  
In particular, 
a general error analysis can be carried out.  
{}From this analysis, 
one concludes that, 
for sufficiently large volumes, 
a limiting $\theta$ exists, 
beyond which reliable measurements 
of the free energy cannot be made.    

For a fixed $V$, 
let $P(Q)$ be the probability of having 
a configuration with topological charge $Q$ 
in some system.  
Let $P_{MC}(Q)$ be the corresponding quantity 
as measured in a Monte Carlo simulation.  
Assume that the Monte Carlo updating procedure  
generates configurations proportional 
to Boltzmann weights.  
Below, this assumption is relaxed.    
If $N_{MC}(Q)$ is the number of times that 
configurations with topological charge $Q$ 
arise in such a simulation, 
then 
\be
  P_{MC}(Q) \equiv  
  { { N_{MC}(Q) } \over { \sum_{Q^\prime} N_{MC}(Q^\prime) } }
\quad . 
\label{eq2p1} 
\ee 
In a typical simulation, 
the measured $P_{MC}(Q)$ differ from the exact $P(Q)$ 
by small errors $\delta P(Q)$: 
\be
  P_{MC}(Q) = P(Q) + \delta P(Q)
\quad . 
\label{eq2p2} 
\ee 
With enough measurements, 
$| \delta P(Q) | \ll 1$.  
In most systems and the ones considered in this work, 
$P(Q)$ falls off%
{\footnote{Eventually the falloff is rapid.}} 
with $Q$, 
so that $P(Q) > P(Q^{\prime})$ 
for $ |Q| < |Q^{\prime} |$.  
A criterion for a simulation to have good statistics 
is that   
$\delta P(Q) \ll P(0)$.  

Let $f( \theta )$ be the difference 
between the free energy ${\cal F} ( \theta )$ 
of a system with a $\theta$ term 
and the free energy of a system with $\theta = 0$: 
\be
  f( \theta ) = {\cal F} ( \theta ) - {\cal F} ( 0 )
\quad . 
\label{eq2p3} 
\ee 
Typically, 
$f( \theta )$ is an increasing function of $\theta$ 
for $0 \le \theta \le \pi$.  
The free energy difference $f( \theta )$ 
is constructed from $P(Q)$ 
using   
\be
  \exp {( - V f( \theta ) )} = 
   \sum_{Q} P(Q) \exp {( i \theta Q )} 
\quad . 
\label{eq2p4} 
\ee 
Normally  
$P(-Q) = P(Q)$,  
so that $f( -\theta ) = f( \theta )$.  

In a Monte Carlo simulation, 
an approximation $f_{MC}( \theta )$ 
to $f( \theta )$ is obtained 
by using $P_{MC}(Q)$ in lieu of $P(Q)$: 
\be
  \exp {( - V f_{MC}( \theta ) )} = 
   \sum_{Q} P_{MC}(Q) \exp {( i \theta Q )} = 
   \exp {( - V f( \theta ) )} + \delta Z( \theta ) 
     \quad ,  
\label{eq2p5} 
\ee 
where 
\be
  \delta Z( \theta ) = 
   \sum_{Q} \delta P(Q) \exp {( i \theta Q )}
\quad .    
\label{eq2p6} 
\ee 
Hence, 
\be
  - V f_{MC}( \theta ) ) = 
  \log { \left[ { \exp {( - V f( \theta ) )} + 
            \delta Z( \theta ) } \right] } 
\quad .    
\label{eq2p7} 
\ee 
Since $f ( \theta ) $ is an increasing function of $\theta$, 
an accurate measurement of $f ( \theta )$ 
for $0 \le \theta < \theta_0$ 
is obtained
if 
\be 
  | \delta Z( \theta ) | \ll  \exp {( - V f( \theta_0 ) )} 
\quad .       
\label{goodcrit} 
\ee 
In particular, 
since $f(0) = 0$ and 
$| \delta Z( \theta ) | \ll 1$, 
there is always a region near $\theta = 0$  
for which $f( \theta) $ can be measured 
in a Monte Carlo simulation.  
However, away from $\theta = 0$, 
Eq.\ \bref{goodcrit} implies that
accurate results are obtained  
only if the error $\delta Z( \theta )$ 
is exponentially small in $V$.    
If Eq.\ \bref{goodcrit} 
is satisfied 
with $\theta_0 = \pi$, 
then a reliable measurement of $f( \theta )$ 
can be made throughout the entire fundamental region 
$0 \le \theta \le \pi$.  

If the inequality 
in Eq.\ \bref{goodcrit} 
is not satisfied for some $\theta_0$, 
then one of several possibilities may arise.  
If, on one hand, 
$\exp {( - V f( \theta_0 ) )} + \delta Z( \theta_0 ) < 0$ 
then the argument in the $\log$ on the right-hand-side 
of Eq.\ \bref{eq2p7} 
becomes negative 
and one will not be able to extract $f( \theta_0 )$ 
from the measurements of the probabilities $P_{MC}(Q)$.  
In Monte Carlo simulations for such a situation, 
a growth in the errors of 
$f_{MC} ( \theta )$ will be observed  
as $\theta$ approaches $\theta_0$,  
and $f_{MC} ( \theta )$ will eventually not be measurable.  
If, on the other hand,  
$\exp {( - V f( \theta_0 ) )} + \delta Z( \theta_0 ) > 0$ 
then  $f_{MC} ( \theta )$ will be measurable 
at $\theta = \theta_0$ 
but the results will not be accurate.  

For sufficiently large $V$, 
a value of $\theta$ exists 
beyond which 
it is impossible to reliable compute $f (\theta )$.  
If we call this value $\theta_b$, 
then $\theta_b$ is the maximum value of $\theta_0$ 
for which 
Eq.\ \bref{goodcrit} 
is satisfied.  
The value of $\theta_b$ 
depends on the statistical accuracy of the simulation.  
As $V$ gets larger, 
$\theta_b$ decreases  
unless an enormously large number of measurements  
are undertaken to reduce statistical errors.  
For large $V$, 
obtaining enough measurements becomes, 
in any practical sense, 
impossible.   
Clearly, 
it is more difficult to measure $f( \theta )$ 
throughout the entire fundamental region of $\theta$, 
as $V$ gets larger.  

It turns out that in most Monte Carlo simulations, 
there is a tendency for 
\be
  | \delta P(0) | > | \delta P(1) | > | \delta P(2) | > \dots 
\quad .    
\label{pineq} 
\ee 
The reason 
for Eq.\ \bref{pineq} 
is explained in the next paragraph.  
For the sake of argument, 
suppose that $| \delta P(0) |$ is much larger than  
$| \delta P(Q) |$ for $| Q | \ge 1$.    
Then, 
from Eq.\ \bref{goodcrit}, 
one concludes that 
\be
   f ( \theta_b ) \approx { {1} \over {V} } | \log | \delta P(0) | | 
\quad .    
\label{estthetab} 
\ee 
Since Monte Carlo results are reliable for $\theta < \theta_b$,  
\be
   f_{MC} ( \theta ) \approx  f ( \theta ) 
\quad {\rm for} \ \theta < \theta_b
\quad . 
\label{eq2p11} 
\ee 
If, in addition, $\delta P(0) > 0$, 
then  
one will find%
{\footnote{ 
Note that since $\delta P(0) \ll 1$, 
the right-hand side of 
Eq.\ \bref{eq2p12} 
is positive.}} 
\be
   f_{MC} ( \theta ) \approx 
   - { {1} \over {V} }  \log \delta P(0)   
\quad {\rm for} \ \theta > \theta_b
\quad ,  
\label{eq2p12} 
\ee 
so that a constant ``flat'' behavior 
in $f_{MC} ( \theta )$ will be observed. 
Although one might expect the statistical error 
in $f_{MC} ( \theta )$ 
to be the order of $f_{MC} ( \theta )$ 
for $\theta \gsim \theta_b$, 
we have observed 
in some simulations that the error, 
as computed by the jackknife method, 
remains anomalously small.  
This happens in both the two-dimensional $U(1)$ 
gauge theory (see Section III) 
and the $CP^{3}$ model (see Section IV).  
When $\delta P(0) > 0$, 
$f_{MC} ( \theta )$ will smoothly interpolate 
between the behaviors in 
Eqs. \bref{eq2p11} and \bref{eq2p12} 
in the region around $\theta \sim \theta_b$. 
The net result is a form for the free energy which resembles
Eq.\ \bref{eq1}. 
If an error $\delta P(Q)$ for $|Q| > 0$ 
is almost as large as $\delta P(0)$, 
then the constant behavior 
in Eq.\ \bref{eq2p12} is replaced by 
a ``slightly wavy'' almost flat curve.  
The above discussion applied to the case $\delta P(0) > 0$. 
If, on the other hand, $\delta P(0) < 0$, 
then it will be impossible to measure $f (\theta )$ 
for $\theta > \theta_b$.  

There are various ways to measure $P_{MC} (Q)$.  
The naive method 
\ct{brsw84} 
is to generate configurations 
proportional to their Boltzmann factor   
and simply count the number $N_{MC} (Q)$ 
of configurations with topological charge $Q$, 
and then use Eq.\ \bref{eq2p1}.  
Since the statistical uncertainty 
$\delta N_{MC} (Q)$ in $N_{MC} (Q)$ 
is proportional to $\sqrt{N_{MC} (Q)}$, 
\be
   | \delta P_{MC} (Q) | \approx c \sqrt{P_{MC} (Q)}
\quad ,     
\label{eq2p13} 
\ee 
for some constant $c$.%
{\footnote{ 
There might be a weak dependence of $c$ on $Q$.}} 
Because $P(Q)$ is a monotonically decreasing function of $Q$ 
for $Q \ge 0$, 
the inequalities 
in Eq.\ \bref{pineq} 
tend to be satisfied.  
In any particular simulation, 
however, 
statistical fluctuations may violate 
Eq.\ \bref{pineq}, 
particularly if $P(Q)$ is only slightly larger 
than $P(Q+1)$.%
{\footnote {
This tends to happen when the volume is large 
and for $Q$ small.}}   

One commonly used Monte Carlo technique 
\ct{bbcs87,wiese89}  
is to perform simulations in $Q$-bins.  
The topological charges $Q$ 
are restricted to be in a bin 
of a certain bin-size.  
The bins must overlap -- usually one takes 
the bins to overlap for only one $Q$ value.    
For example, 
the bins can be taken to be $0 \le Q \le \Delta Q$, 
$\Delta Q \le Q \le 2 \Delta Q$,
$2 \Delta Q \le Q \le 3 \Delta Q$, etc., 
where $\Delta Q$ is a small positive integer. 
Here, the bin-size is $\Delta Q +1$.    
The relative probabilities within a $Q$-bin 
are measured 
by generating configurations which are proportional 
to their Boltzmann factors 
but with their topological charges contained  
in a $Q$-bin.  
Overall probabilities are determined 
from the relative probabilities 
by matching results at the overlap endpoints of two $Q$-bins.  
Compared to the naive method, 
one generates more configurations with large $Q$ configurations.  
Hence, errors in $\delta P(Q)$ are relatively reduced 
for large $Q$.  
This tends to enhance the inequalities 
in Eq.\ \bref{pineq}.  

A benefit of the binning technique 
is that it enables $P(Q)$ to be accurately measured 
for large $Q$, 
even where $P(Q)$ is extremely small.  
The effect can be spectacular: often $P(Q)$ which are 
orders and orders of magnitude smaller that $P(0)$ 
can be measured.  

Another simulation method 
introduces a weight factor $w(Q)$. 
One then generates Monte Carlo configurations 
which are proportional to 
$(the \ Boltzmann \ factor) \times w(Q)$.  
If $\tilde N_{MC}(Q)$ 
is the number of configurations with topological charge $Q$ 
which are generated by such a procedure, 
then Eq.\ \bref{eq2p1} 
is replaced by 
\be
  P_{MC}(Q) \equiv  
  { { \tilde N_{MC}(Q) \left[ w(Q) \right]^{-1} } 
    \over { \sum_{Q^\prime}\tilde N_{MC}(Q^\prime) 
        \left[ w (Q^\prime) \right]^{-1} } }
\quad . 
\label{eq2p14} 
\ee 
Although any weight $w(Q)$ may be used, 
the trial-probability-distribution method 
tries to choose $w(Q)$ 
so that all $Q$ sectors are ``visited'' roughly 
the same number of times.\ct{ksc88} 
In other words, 
one guesses a trial probability 
$P_0 (Q)$, 
which might accurately approximate 
the true probability distribution $P(Q)$.    
One then uses $w(Q) = 1 / P_0 (Q)$.  
If one were to pick $P_0$ so that $P_0 (Q) = P(Q)$, 
then a constant distribution in $Q$ for $\tilde N_{MC} (Q)$ 
would be generated (up to statistical fluctuations) 
and Eq.\ \bref{eq2p14} 
would lead to $P_{MC}(Q) = P_0 (Q) = P(Q) $.  
Like the binning method, 
the trial-probability-distribution method 
generates more configurations 
in the large $Q$ sectors, 
thereby enabling one to better measure $P(Q)$ for $Q$ large.  
This method also enhances the inequalities 
in Eq.\ \bref{pineq}.  

Finally, one can combine the above two methods 
by using a trial probability distribution $P_0 (Q)$ 
with a  $Q$-binning.  
This trial-probability-binning method is excellent 
for measuring $P(Q)$ for large $Q$.  
However, of the four methods mentioned here, 
it enhances the inequalities 
in Eq.\ \bref{pineq} 
the most.  

In summary, 
current methods of simulating systems with $\theta$ terms 
tend to generate errors for $P(Q)$ 
which are ordered as 
in Eq.\ \bref{pineq}.  

None of the methods solve 
the $\theta_b$ barrier problem 
discussed above between 
Eqs.\ \bref{goodcrit} and \bref{eq2p13}:  
Assuming that $\delta P(0)$ dominates 
in the error $\delta Z (\theta)$, 
one sees that 
increasing the statistics hardly changes $\theta_b$ 
because of the logarithm and the volume factor 
in Eq.\ \bref{estthetab}.  
The same statement holds 
even when $\delta P(0)$ is not the dominant error.  
One concludes that 
all simulation methods must fail 
for sufficiently large $\theta$,  
if $V$ of the system is large. 
Cluster algorithms might help a little in this regard. 
\ct{bpw95} 

{}From the above discussions, 
the following general guidelines 
concerning Monte Carlo simulations of systems 
with a $\theta$ term are obtained:    

\begin{itemize}
\item[(1)] When the volume is sufficiently big, 
a limiting $\theta_b$ arises.  
For $\theta > \theta_b$, 
the free energy cannot be reliably measured.  
\item[(2)] As long as finite-size effects are under control, 
that is $\xi < V^{(1/d)}$, 
small-volume results for the measurement of $f (\theta )$ 
are more reliable than large-volume results. 
Here,   
$\xi$ is the correlation length 
and $d$ is the number of dimensions of the system.  
\item[(3)] If a flattening behavior of the free energy $f( \theta)$ 
for large $\theta$ is observed, 
one should be cautious that the result is spurious.  
In particular, 
one should try to see whether $| \delta P(0) |$ 
is bigger than the other $| \delta P(Q) |$.  
\item[(4)] When $P_{MC}(Q)$ is less than $| \delta P(0) |$, 
the contribution to $P_{MC}(Q)$ 
in Eq.\ \bref{eq2p5} 
need not be included.   
The reason for this statement 
is that the contribution of $P_{MC}(Q)$ 
is lost in the ``noise'' 
of the error term $\delta Z$ 
in Eq.\ \bref{eq2p5}.  
\item[(5)] Although the $\theta_b$ barrier cannot be overcome, 
Monte Carlo procedures should emphasize  
accurate measurements of $P(Q)$ for $Q$ near $0$.  
In particular, 
the optimal procedure is one 
for which all $| \delta P(Q) |$ are approximately equal.  
This minimizes the chances 
for anomalous flat behavior in $f( \theta )$.  
\end{itemize}
Combining (2) and (3), 
one obtains another guideline: 
\begin{itemize}
\item[(6)] If a large-volume simulation shows a flattening effect 
for $f( \theta)$ for $\theta$ sufficiently large, 
but a smaller-volume simulation does not, 
one should probably trust the smaller-volume result.  
\end{itemize}

Surprisingly, 
many previous Monte Carlo studies 
of systems with $\theta$ terms 
measure $P(Q)$ over many orders of magnitude.  
Point (4) above says that, although this is not harmful, 
it is an inefficient use of computer time 
if one is interested in measuring the free energy.  
Likewise, 
point (5) implies that the standard binning 
and trial-probability methods 
are not optimal 
because they enhance the inequalities  
in Eq.\ \bref{pineq}.  
The binning method can be improved  
by doing more measurements in the $Q$-bin containing 
$Q=0$ and less measurements in large-$Q$ $Q$-bins.  
Another improvement is as follows. 
One can adjust $w(Q)$ in 
Eq.\ \bref{eq2p14} 
so that the errors $| \delta P(Q) |$ 
are equal (up to statistical effects).  
It is not hard to show 
that the optimal $w(Q)$ is $P(Q)$. 
This corresponds to using a trial-probability function 
which is the inverse of the true probability.    
In other words, the optimal weighting method 
is the antithesis of the trial-probability method 
and the antithesis of 
what is commonly employed in Monte Carlo experiments.

\bigskip 
{\bf\large \noindent III.\ The Lattice  $U(1)$ Gauge Theory}  
\setcounter{section}{3}   
\setcounter{equation}{0}   

The two-dimensional lattice $U(1)$ gauge theory 
serves an important test case because 
computer simulations can be compared 
to the exact analytic results below.  
For related Monte Carlo investigations, 
see refs.\ \ct{wiese89,panagiotakopoulos85}.  

In the lattice formulation of a gauge theory, 
one assigns an element of the group to each link of the lattice. 
For the $U(1)$ case, 
such an element is a phase.  
The lattice $U(1)$ gauge theory action is 
\be
    S^{U(1)} = \beta \sum_p 
             \left( { U_p + U_p^{*}  } \right) 
\quad , 
\label{eq3p1}
\ee 
where $U_p$ is the product of the $U(1)$ link phases 
around the plaquette $p$ 
and where $\beta$ is the inverse coupling.  

Define a local topological density $\nu_p $ via  
$\nu_p \equiv \log \left( { U_p } \right) / (2\pi)$, 
where $ -\pi < \log \left( { U_p } \right) \le \pi$.  
The total topological charge $Q$ is given by 
$Q = \sum_p \nu_p$.  
The theta term $S_{\theta \ {\rm term}}$ is $i \theta Q$,  
that is, \ct{bl81} 
\be 
  S_{\theta \ {\rm term}} = 
  {{ i\theta } \over {2\pi} } \sum_p 
   \log \left( { U_p } \right) 
\quad . 
\label{eq3p2} 
\ee 
Eq.\ \bref{eq3p2} is the lattice analog 
of the continuum $\theta$-term action 
$i { {\theta} \over {2 \pi} } \int d^2 x F$. 

In $d=2$ dimensions, 
gauge theories are exactly solvable 
even in the presence 
of a $\theta$ term.  
For periodic boundary conditions, 
in which case $Q$ is quantized to an integer, 
the result is 
\ct{wiese89}  
\be 
  Z \left( {\theta , \beta, V } \right) = 
    \sum_{m = -\infty}^{\infty} 
           \left[ { {z} \left( {\theta + 2 \pi m, \beta } \right) 
                  } \right)]^V 
\quad ,  
\label{eq3p3} 
\ee 
where 
\be 
   {z} \left(  { \theta , \beta } \right)  = 
  { { \int_{-\pi}^{\pi} {{df_p} \over {2\pi}} 
      \exp ({ {i f_p \theta} \over {2\pi} }) 
         \exp ( 2\beta \cos (f_p) )  } 
       \over 
    { \int_{-\pi}^{\pi} {{df_p} \over {2\pi}}  \exp ( 2\beta \cos (f_p) ) } 
  }
\quad .   
\label{eq3p4} 
\ee 
Here, $f_p$ can be thought of as the field strength 
for a single plaquette: $U_p =  \exp ( i f_p )$.   
In the infinite volume limit, 
the free-energy difference per unit volume $f (\theta) $ is given by 
\be 
   f (\theta , \beta ) = 
     - \log \left( { {z} \left(  {\theta , \beta } \right) 
                   } \right)
 \quad .   
\label{eq3p5} 
\ee 
In particular, 
at $\beta = 0$, 
one obtains \ct{seiberg84}
\be 
   f (\theta , 0 ) = 
     - \log ( {{2} \over {\theta}} \sin( {{\theta} \over {2}} ) ) 
 \quad ,    
\label{eq3p6} 
\ee 
when $-\pi < \theta \le \pi$.  

We have performed Monte Carlo studies 
of the $U(1)$ gauge theory 
to gain insight into computer simulations 
for a system with a $\theta$ term.  
The action consisted of 
the sum of the actions 
in Eqs.\ \bref{eq3p1} and \bref{eq3p2}.  
Two values of $\beta$ were considered: 
$\beta = 0.0$ and $\beta = 1.0$.  
Three simulation methods were employed: 
naive, binning and binning with a trial probability function.  
Heat-bath updating was used 
with the naive and binning methods. 
The Metropolis algorithm was used 
with the trial-probability-binning method.  
The trial probability distribution $P_0 (Q)$ 
was chosen to be a gaussian: 
$ P_0 (Q) \propto \exp( - k  Q^2)$, 
where the constant $k$ was appropriately selected.
After thermalizing the system, 
the number of sweeps  
ranged from tens of million to several hundred million.  
One sweep corresponded to updating once all the link-variables 
of the lattice. 
The number-crunching  
was done on IBM and SUN desktop workstations.  

\begin{figure}[t]
\epsfxsize=5in
\epsfbox{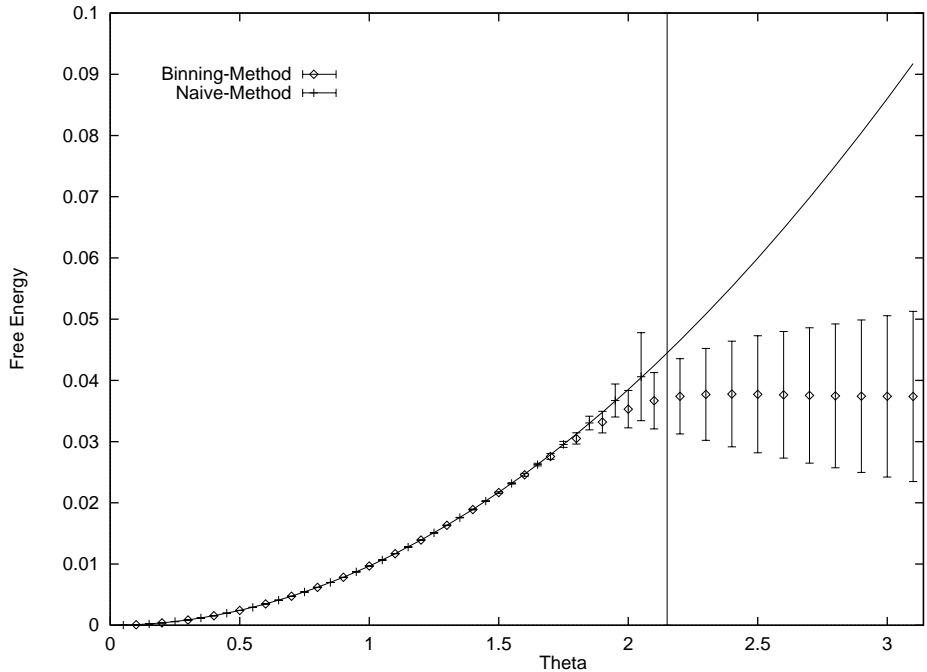}
\caption{$U(1)$ Free Energy Versus $\theta$ at $\beta=1.0$ 
for the Naive and Binning Methods.}
\end{figure}

Figure 1 plots the free energy versus $\theta$ 
for $\beta = 1.0$, 
for a periodic $16 \times 16$ lattice.  
For the naive method, 
the data points correspond to 
short horizontal line segments.   
A total of $75$ million updating sweeps were performed.  
The error bars 
were computed using a jackknife method 
\ct{jackknife} 
by dividing the run into $15$ data sets,  
each of which involved $5$ million sweeps. 
The solid line is the exact analytic result.  
Analytic and Monte Carlo results agree 
for $\theta$ less than $2.1$.  
The agreement, 
which is excellent, 
cannot be seen on the scale of Figure 1.  
For example, 
at $\theta = 0.5$, 
$f_{MC} = 0.00242004 (53)$ 
versus 
$f_{\rm exact} = 0.0024196$, 
at $\theta = 1.0$, 
$f_{MC} = 0.0096685 (46)$ 
versus 
$f_{\rm exact} = 0.0096682$, 
at $\theta = 1.5$, 
$f_{MC} = 0.021699 (61)$ 
versus 
$f_{\rm exact} = 0.021714$, 
and 
at $\theta = 2.0$, 
$f_{MC} = 0.0386 (37)$ 
versus 
$f_{\rm exact} = 0.0385$, 
where the statistical uncertainty 
in the last two digits is displayed in parenthesis.%
{\footnote{For example, 
the $\theta = 2.0$ results is 
$f_{MC} = 0.0386 \pm 0.0037$.}} 
For $\theta$ beyond $2.1$, 
error bars grew and the partition function 
became negative.  
One sees that the ``barrier theta'' 
$\theta_b$ is about $2.1$.  
The statistical error in $ P_{MC}(0)$ 
was $3 \times 10^{-5}$. 
Using this error in 
Eq.\ \bref{estthetab} 
to estimate $\theta_b$,  
one finds $\theta_b \approx 2.05$. 
The agreement of the theoretical $\theta_b$ 
with the Monte Carlo value confirms the data-analysis 
discussion of Section II. 
One can also check Eq.\ \bref{eq2p13} 
by seeing whether the ratio 
of the statistical error 
of $ P(Q)$ 
to $\sqrt { P_{MC}(Q) }$ 
remains constant.  
Even though $P_{MC}(Q)$ 
varied by eight orders of magnitude, 
it turned out that the ratio stayed constant 
to within a factor of $3$.  

Figure 1 also plots the free energy versus $\theta$ 
and for the binning method,  
again 
for $\beta = 1.0$  
on a $16 \times 16$ lattice. 
The data points correspond to diamonds.    
Five bins of bin-size $4$ were used.  
For each bin, $16$ million sweeps were performed.  
Since $5$ bins were used, 
the statistics for this case 
are comparable to the statistics 
of the naive method of the previous paragraph.  
The $16$ million sweeps were divided into $20$ sets 
for the jackknife analysis of errors.  
For $\theta \ge 2.0$, 
the Monte Carlo data for the free energy 
dropped below the exact result 
and became constant. 
For $\theta \le 1.7$, 
the agreement between Monte Carlo data 
and the analytic result 
was comparable to the naive-method case, 
discussed in the previous paragraph. 
As one can see in Figure 1, 
the flat behavior  
in the free energy is anomalous 
for $\theta > 2.1$ 
even though the error bars are sizeable.  
At $\theta = \pi$, 
the discrepancy with the exact analytic result 
for $f$ 
is at the $4 \sigma$ level. 

\begin{table}[t]
\begin{center}
\begin{tabular}{|r|r|r|r|r|r|r|}
\cline{1-3} \cline{5-7}\cline{1-3} \cline{5-7}
    $Q$  &\multicolumn{0}{c|}{$P_{MC}(Q)$}&
  \multicolumn{0}{c|}{$P_{\mbox{exact}}(Q)$} 
& & $Q$  & \multicolumn{0}{c|}{$P_{MC}(Q)$}  
  &\multicolumn{0}{c|}{$P_{\mbox{exact}}(Q)$}  
\\
\cline{1-3} \cline{5-7}
 0 &  1.968226(42)\, E-01 &     1.968199\, E-01 &&
 7 & 4.51534(28)\, E-04 &  4.51495\, E-04 \\
 1 & 1.743842(23)\, E-01  &   1.743866\, E-01 &&
 8 & 6.58573(46)\, E-05  &  6.58533\, E-05  \\
 2 & 1.212054(16)\, E-01  &    1.212086\, E-01 &&
 9 & 7.13760(58)\, E-06  &  7.13774\, E-06 \\
 3 & 6.59478(17)\, E-02  &    6.59454\, E-02 &&
 10 & 5.64442(51)\, E-07 &   5.64463\, E-07 \\
 4 & 2.79818(09)\, E-02   &    2.79806\, E-02 &&
 11 & 3.18299(30)\, E-08 &   3.18344\, E-08 \\
 5 & 9.20984(42)\, E-03  &  9.20931\, E-03 &&
 12 & 1.24331(15)\, E-09 &  1.24034\, E-09 \\
 6 & 2.33452(13)\, E-03 &  2.33437\, E-03 && & & \\
\cline{1-3} \cline{5-7}
\end{tabular}
\end{center}
\caption{$P(Q)$ for $U(1)$ Model at $\beta=1.0$ on a $7\times 7$ Lattice.}
\end{table}

We also performed a simulation 
with a trial-probability-binning method, 
again 
with $\beta = 1.0$ on a $16 \times 16$ lattice.  
The results were similar quantitatively 
to the naive case above, 
except that the Monte Carlo data dropped 
below the exact result 
at about $2.1$ and then the error bars became enormous 
for $\theta \ge 2.2$.  

In all three of the above runs, 
the statistical errors in $ P(Q)$  
were ordered as 
in Eq.\ \bref{pineq}.  

For the binning-method case, 
one can address the question 
of whether the flattening behavior 
is attributable to a value of $P_{MC}(0)$ 
which is larger than $P(0)$. 
For a $16 \times  16$ lattice, 
the exact $P(0)$ is $0.179259$.  
The value of $P_{MC}(0)$ for the binning run 
was $0.179295 (72)$.  
Thus, 
$P_{MC}(0)$ was indeed greater than $P(0)$.  
For the naive run, 
$P_{MC}(0) = 0.179242 (30)$, 
while for the trial-probability-binning run, 
$P_{MC}(0) = 0.179192 (55)$, 
and   
since these two values 
are below the exact $P(0)$, 
the anomalous flattening behavior 
is not expected to arise, 
in agreement with the Monte Carlo results.

According to Eq.\ \bref{estthetab}, 
one can increase $\theta_b$ 
by decreasing the size of the system.  
We therefore estimated that if the lattice size 
was $8$, 
then a reliable measurement of the free energy 
could be made 
throughout the entire fundamental $\theta$ region.  
Using a trial-probability-binning method 
on an $8 \times 8$ lattice, 
reasonable agreement 
between Monte Carlo data and analytic results 
was obtained 
in a $30$-million sweep run. For example, at $\theta=1.0$,
$f_{MC}=0.00980(5)$ versus $f_{\mbox{exact}}=0.00967$, at
$\theta=2.0$, $f_{MC}=0.0382(1)$ versus $f_{\mbox{exact}}=0.0385$
and at $\theta=3.0$,
$f_{MC}=0.089(3)$ versus $f_{\mbox{exact}}=0.0814$.

\begin{table}[t]
\begin{center}
\begin{tabular}{|r|r|r|r|r|r|r|}
\cline{1-3} \cline{5-7}\cline{1-3} \cline{5-7}
    $\theta$  &\multicolumn{0}{c|}{$F_{MC}$}&
  \multicolumn{0}{c|}{$F_{\mbox{exact}}$} 
& & $\theta$  & \multicolumn{0}{c|}{$F_{MC}$}  
  &\multicolumn{0}{c|}{$F_{\mbox{exact}}$}  
\\
\cline{1-3} \cline{5-7}
   0.2 &  1.667250(80)\, E-03  &    1.667220\, E-03 &&
    1.8 &   1.38842(14)\, E-01   &      1.38845\, E-01 \\
   0.4 &   6.67569(29)\, E-03  &    6.67558\, E-03 &&
   2.0 &   1.72596(23)\, E-01   &      1.72603\, E-01 \\
   0.6 &   1.504550(71)\, E-02 &     1.504530\, E-02 &&
    2.2 &   2.10467(38)\, E-01   &      2.10485\, E-01 \\
   0.8  &  2.68107(14)\, E-02  &    2.68104\, E-02 &&
    2.4 &   2.52636(72)\, E-01   &       2.52674\, E-01 \\
    1.0 &   4.20200(25)\, E-02 &     4.20195\, E-02 &&
    2.6 &   2.9917(15)\, E-01    &      2.9925\, E-01 \\
    1.2 &   6.07374(40)\, E-02 &     6.07369\, E-02 &&
    2.8 &   3.4907(35)\, E-01    &      3.4926\, E-01 \\
    1.4 &   8.30438(62)\, E-02  &    8.30436\, E-02 &&
    3.0 &   3.9436(75)\, E-01     &      3.9472\, E-01 \\
    1.6 &   1.090390(94)\, E-01  &       1.090390\, E-01 && & & \\
\cline{1-3} \cline{5-7}
\end{tabular}
\end{center}
\caption{Measured and Exact Free Energy for $\beta=0.0$ on a $4 \times
4$ Lattice.}
\end{table}

Consider now the infinite-strong-coupling case $\beta = 0.0$.  
Results for this case 
were qualitatively similar to the $\beta = 1.0$ case: 
When the lattice size was sufficiently small, 
the free energy was measurable over the entire 
fundamental $\theta$ region 
and the data agreed well with exact analytic calculations.  
When the lattice size was bigger, 
the free energy was accurately measurable 
only for $0 \le \theta < \theta_b$. 
For example, 
a run, which used the naive method on a $30 \times 30$ lattice 
and which involved $22$ million sweeps,  
produced good results only for $0 \le \theta < 0.49$. 
The barrier value was in agreement 
with $\theta_b$ of $0.5$,
as computed 
from Eq.\ \bref{goodcrit}.%
{\footnote 
{On this large lattice $P(Q)$ 
decreased slowly for $Q$ near zero, 
so that the largest statistical error 
actually occurred for $P(-4)$.  
This error was $1.4 \times 10^{-5}$.}}  
When a binning method was used, 
the free-energy data points began to slip below 
the analytic results near $\theta \approx 0.45$. 
A slightly-wavy-but-basically-flat behavior 
for the free energy 
was observed for $0.45  < \theta < 0.80$.  
In this range, 
the Monte Carlo data was below the exact analytic result 
at the $2 \sigma$ level.  
For $\theta > 0.81$, 
the error bars become large 
and the free energy was not measurable.  
When the lattice size was reduced, 
the free energy was measurable 
over a larger $\theta$ region.  
On a $7 \times 7$ lattice, 
we performed a run 
with the trial-probability-binning method.  
For the first $180$ million sweeps, 
a graph qualitatively similar 
to the binning case of Figure 1 
was obtained.  
The free energy was measurable up to $\theta_b \approx 2.1$.  
For $\theta > 2.1$, the data points fell below the exact result 
and a flat free-energy curve was produced.  
{}From the error in $P(0)$ of $\sim 10^{-5}$, 
the theoretically predicted value of $\theta_b$ 
from Eq.\ \bref{estthetab} was $2.3$. 
When the run was continued, 
the flat behavior went away:   
After a billion sweeps, 
the data agreed with analytic results for $ 0 \le \theta < 2.2$. 
Beyond $2.2$, the data fell slightly below the exact results 
and the error bars became large at $\theta = 2.7$. 
For $\theta > 2.7$, 
the free energy was no longer numerically measurable.  
Based on Eq.\ \bref{estthetab}, 
the limiting $\theta_b$ should be $2.65$ 
for this case, 
in agreement with the Monte Carlo results.  
Table 1 displays the kind of accuracy 
with which $P(Q)$ was measured.  
Finally, we reduced the lattice size to $4$. 
In a run using the binning with a probability distribution, 
the free energy was accurately obtained 
throughout the fundamental $\theta$ region.  
A run with $525$ million sweeps was performed 
with $2$ bins of bin-size $4$.  
Table 2 provides the comparison 
of Monte Carlo and exact results for the free energy. 

The $\beta = 0.0$ results of this section 
are relevant for the $CP^{N-1}$ models at $\beta = 0$:
When $\beta = 0$, 
the $CP^{N-1}$ 
models coincide with $U(1)$ gauge theory.  
A numerical investigation of the $CP^{3}$ 
model at non-zero $\beta$
is the subject of Section IV.

\bigskip 
{\bf\large \noindent IV.\ The Lattice $CP^3$ Model}  
\setcounter{section}{4}   
\setcounter{equation}{0}   

Monte Carlo studies of the ``adjoint'' form 
of the lattice $CP^3$ model 
in the presence of a $\theta$ term have been performed 
in ref.\ \ct{schierholz94}.  
For our Monte Carlo investigations, 
we have selected the ``auxiliary $U(1)$ field'' formulation 
for two reasons: 
(1) it allows us to investigate the $CP^3$ model 
from a different-but-equally-good form of the lattice action, 
and more importantly, 
(2) strong coupling expansions 
\ct{ps96a}  
have been obtained 
for this form of the action, 
thereby allowing comparisons of Monte Carlo data 
with analytic results, 
at least for small inverse coupling $\beta$.  
The $CP^{N-1}$ models without a $\theta$ term have been studied 
by computer simulations in several works. 
See refs.\ \ct{hm92}-\ct{vicari93}. 

The lattice $CP^{N-1}$ action, which we employ, is 
\be 
  S = \beta N \sum_{x,\Delta} 
 \left( { 
    z_{x}^\ast \cdot z_{x+\Delta} 
    U \left({  x, x+\Delta } \right) + c.c. 
        } \right) 
\quad , 
\label{eq4p1} 
\ee 
where the complex scalar fields $z_{x}^i$ 
satisfy $\sum_{i=1}^N z_{xi}^\ast z_{x}^i =1 $ 
and where $c.c.$ is the complex conjugate of the first term 
in Eq.\ \bref{eq4p1}. 
Here,   
the sum over $\Delta$ 
involves the $d$ positively-directed nearest neighbors to $x$,  
so that $\Delta$ takes on the values 
$e_1$, $e_2$,$\dots$, $e_d$, 
where $e_i$ is a unit vector in the $i$th direction.  
Since we consider the two-dimensional case, $d=2$. 
The field 
$U \left({  x, x + \Delta } \right)$ 
is a phase associated with link between $x$ and $x + \Delta$  -- 
it is the same link-variable  
which appears in the $U(1)$ gauge theory 
of Section III.  
To  Eq.\ \bref{eq4p1}, 
we add $S_{\theta}$ 
of Eq.\ \bref{eq3p2}  
to obtain the full action.  
Finally, we treat the $N=4$ case, i.e., the $CP^3$ model.  

\begin{figure}[t]
\epsfxsize=5in
\epsfbox{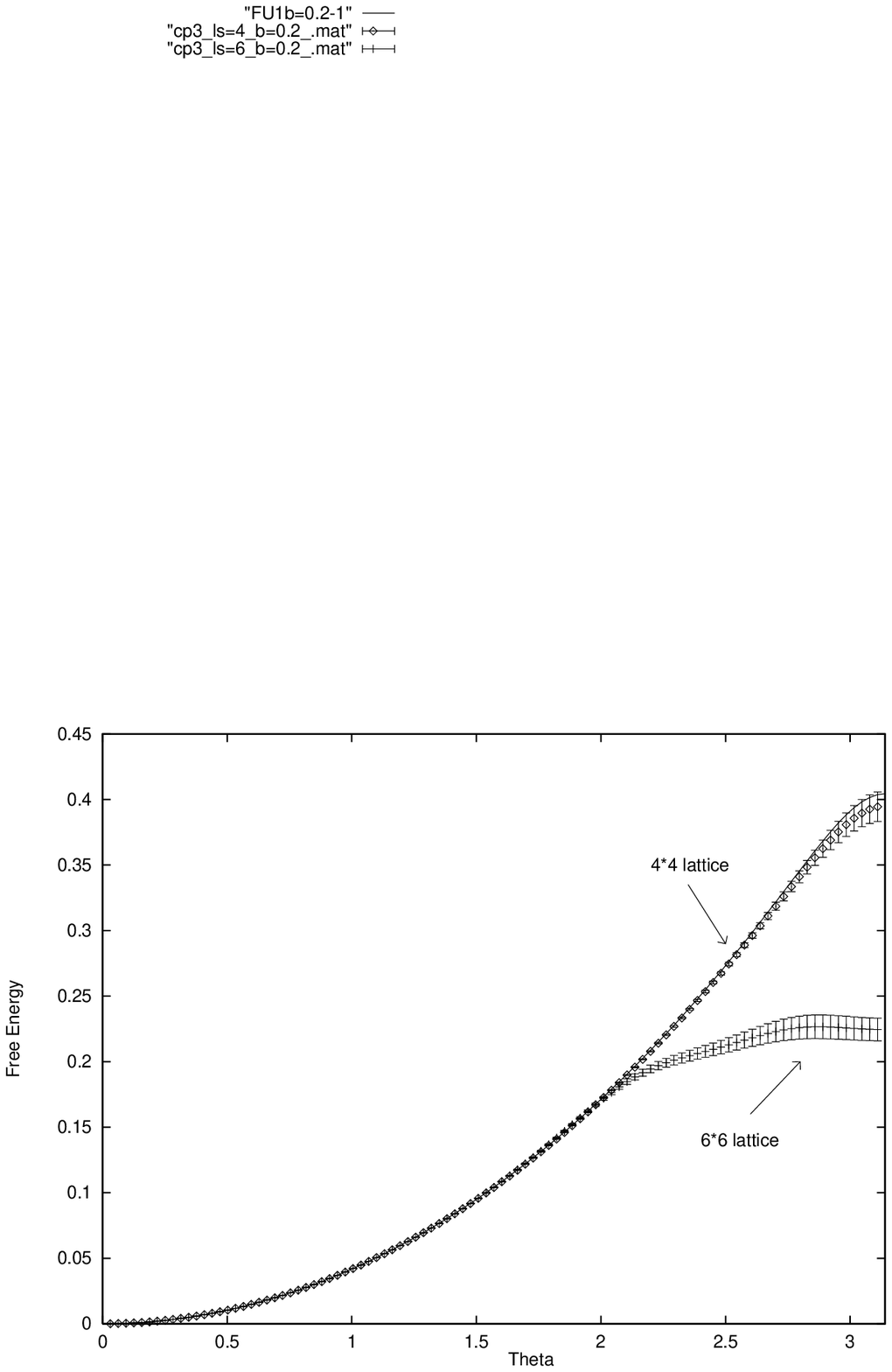}
\caption{$CP^3$ Free Energy Versus $\theta$ at $\beta=0.2$ 
on $4\times 4$ and $6 \times 6$ Lattices.}
\end{figure}

We used the Metropolis algorithm 
for all degrees of freedom. 
The $z_x^i$ fields were separated into $4$ real 
and $4$ imaginary components.  
They were updated by performing rotations 
in each of the $28$ planes 
of the $8$-dimensional real vector space. 
For the $U(1)$ fields, a binning method was used.  
After each time-consuming update of the $z_x^i$ fields, 
$10$ $U(1)$ updates, as well as $Q$ measurements, 
were carried out 
in the fixed $z_x^i$ background.
The total number of $U(1)$ sweeps
ranged from three to $150$ million. 
Simulations were done 
for $\beta = 0.2, 0.6, 0.7, 1.0$ and $1.1$. 
The intermediate-coupling cross-over region  
is around $0.8$, 
so that the latter two $\beta$ values 
are in the weak coupling region 
where continuum scaling should set in.  

In some simulations with large volumes, 
the free energy was not measurable 
beyond a certain value of $\theta$: 
error bars became very large 
or the partition function went negative. 
The value of $\theta$ at which this occurred 
was in approximate agreement with $\theta_b$ 
obtained using 
Eq.\ \bref{estthetab}. 
Below, 
we show results for cases 
in which the free energy 
was measurable throughout the entire fundamental $\theta$ region  
or for cases in which a flat behavior was observed.  

Figure 2 shows the free energy 
at $\beta = 0.2$ 
for $4 \times 4$ and $6 \times 6$ lattices. 
The solid line represents the tenth-order strong-coupling 
character expansion of ref.\ \ct{ps96a}. 
At this small value of $\beta$, 
the strong-coupling expansion 
should be quite close to the exact result.  
On the $6 \times 6$ lattice, 
an anomalous flattening behavior was observed.  
The discrepancy between Monte Carlo data and 
the strong-coupling series was more than $10 \sigma$ 
for $\theta$ near $\pi$.  
For some reason, 
the jackknife error analysis 
produced error bars which did not come close 
to overlapping with the true results.  
Furthermore, 
Eq.\ \bref{estthetab} 
predicts that $\theta_b$ should be $2.1$, 
so that measurements of the free energy should not 
be reliable for $\theta > 2.1$. 
This theoretical estimate for $\theta_b$ 
is close to the point 
where constant free-energy behavior set in.  
When the lattice size was reduced to $4$, 
agreement between Monte Carlo data and 
the strong-coupling series occurred 
throughout the entire fundamental $\theta$ region.  

\begin{figure}[t]
\epsfxsize=5in
\epsfbox{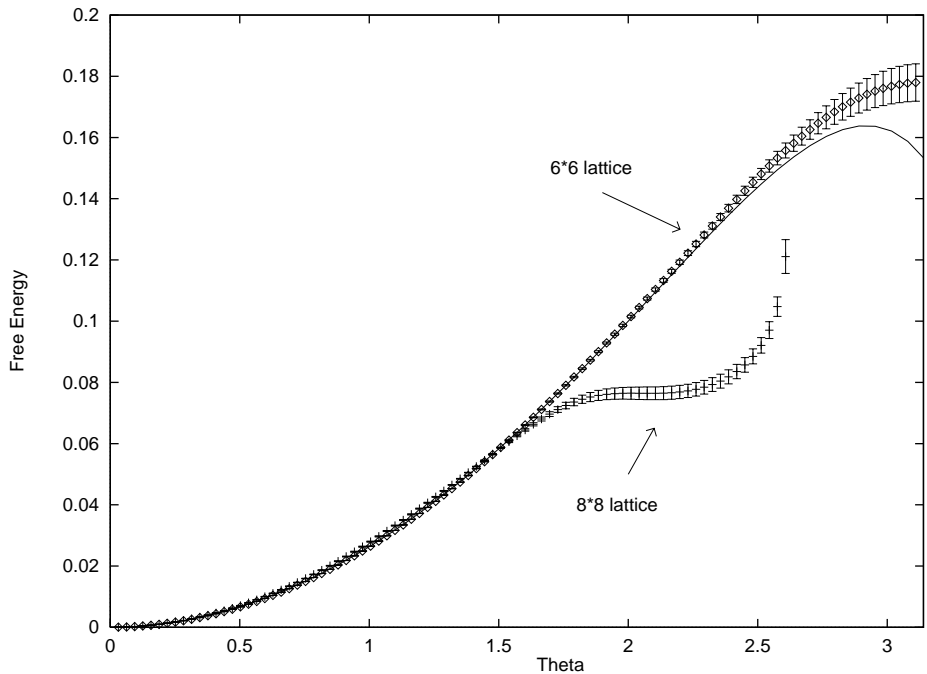}
\caption{$CP^3$ Free Energy Versus $\theta$ at $\beta=0.6$ 
on $6\times 6$ and $8\times 8$ Lattices.}
\end{figure}

At $\beta = 0.6$, 
roughly similar behaviors in the Monte Carlo results 
were obtained 
except, that on the $8 \times 8$ lattice, 
a flat behavior in the free energy began at $\theta = 1.7$, 
which turned 
up at $\theta = 2.5$.   
The free energy was not measurable beyond $\theta = 2.6$.  
See Figure 3.  
Eq.\ \bref{estthetab} predicts 
that free energy Monte Carlo results 
should not be trusted for $\theta > 1.75$.  
For a $6 \times 6$ lattice, 
agreement with the strong-coupling series 
was obtained up to about $\theta = 2.5$.  
Beyond that point, 
the strong-coupling series was slightly below Monte data 
(at roughly the $1 \sigma$ level).  
At $\theta = 2.8$, 
the strong-coupling expansion of the free energy peaks 
and for larger $\theta$ it decreases, 
while Monte Carlo data continued to increase.   

\begin{figure}[t]
\epsfxsize=5in
\epsfbox{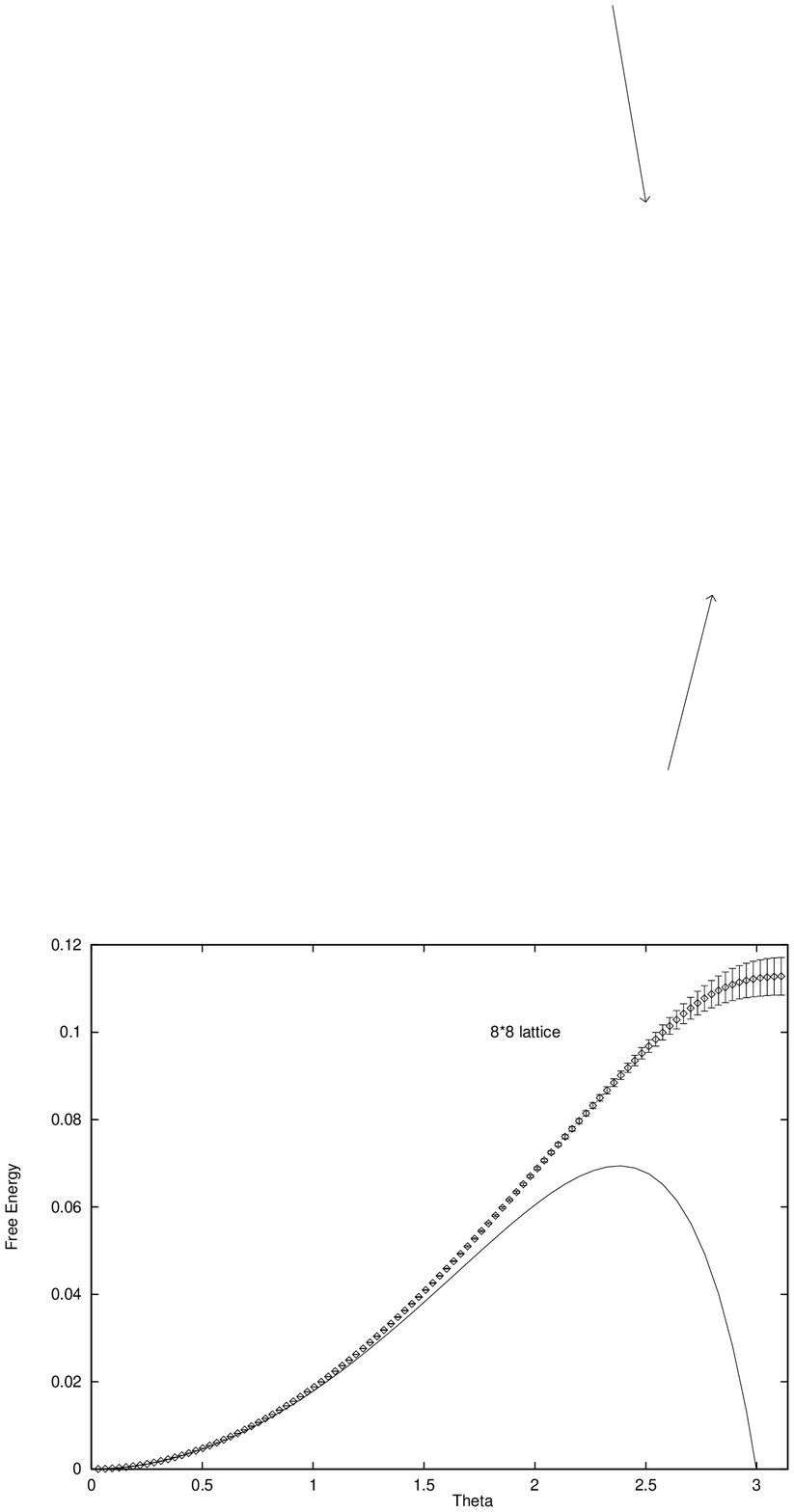}
\caption{$CP^3$ Free Energy Versus $\theta$ at $\beta=0.7$ 
on a $8\times 8$ Lattice.}
\end{figure}

At $\beta = 0.7$, 
the strong-coupling series peaks at $\theta = 2.4$ 
and then decreases significantly.  
Since the free-energy should be a non-decreasing function 
of $\theta$, 
higher-order-in-$\beta$ corrections 
are probably important in this region, 
as ref.\ \ct{ps96a} suggested might be the case.  
The Monte Carlo data began to rise 
above the strong-coupling results 
at around $\theta = 1.2$.  
See Figure 4.  
At this value of $\beta$, 
which is in the intermediate coupling region, 
the Monte Carlo data 
is to be trusted over 
the strong-coupling expansion.  

At $\beta = 1.0$,  
which is on the weak coupling side 
of the intermediate coupling region,
a simulation on a $20 \times 20$ lattice 
gave good results throughout the fundamental $\theta$ region.  
Figure 5 displays the free energy versus $\theta$.  
The solid curve is the fit 
$f( \theta ) = 0.0039 * ( 1 - \cos ( \theta ) )$.  
It reproduces the data within error bars.  
At $\beta = 1.1$, 
a simulation on a $40 \times 40$ lattice 
also gave good results.  
The free energy function 
$f( \theta ) = 0.00125 * ( 1 - \cos ( \theta ) )$
reproduces the $\beta = 1.1$ data well 
throughout the fundamental $\theta$ region. 
It is interesting that a ``cosine'' form 
fits the data for $\beta \ge 1.0$.  
Such a functional form arises 
from a topological gas picture \ct{cdg76}.  
For $\beta < 0.8$, 
a cosine form did not fit the free energy data.  

\begin{figure}[t]
\epsfxsize=5in
\epsfbox{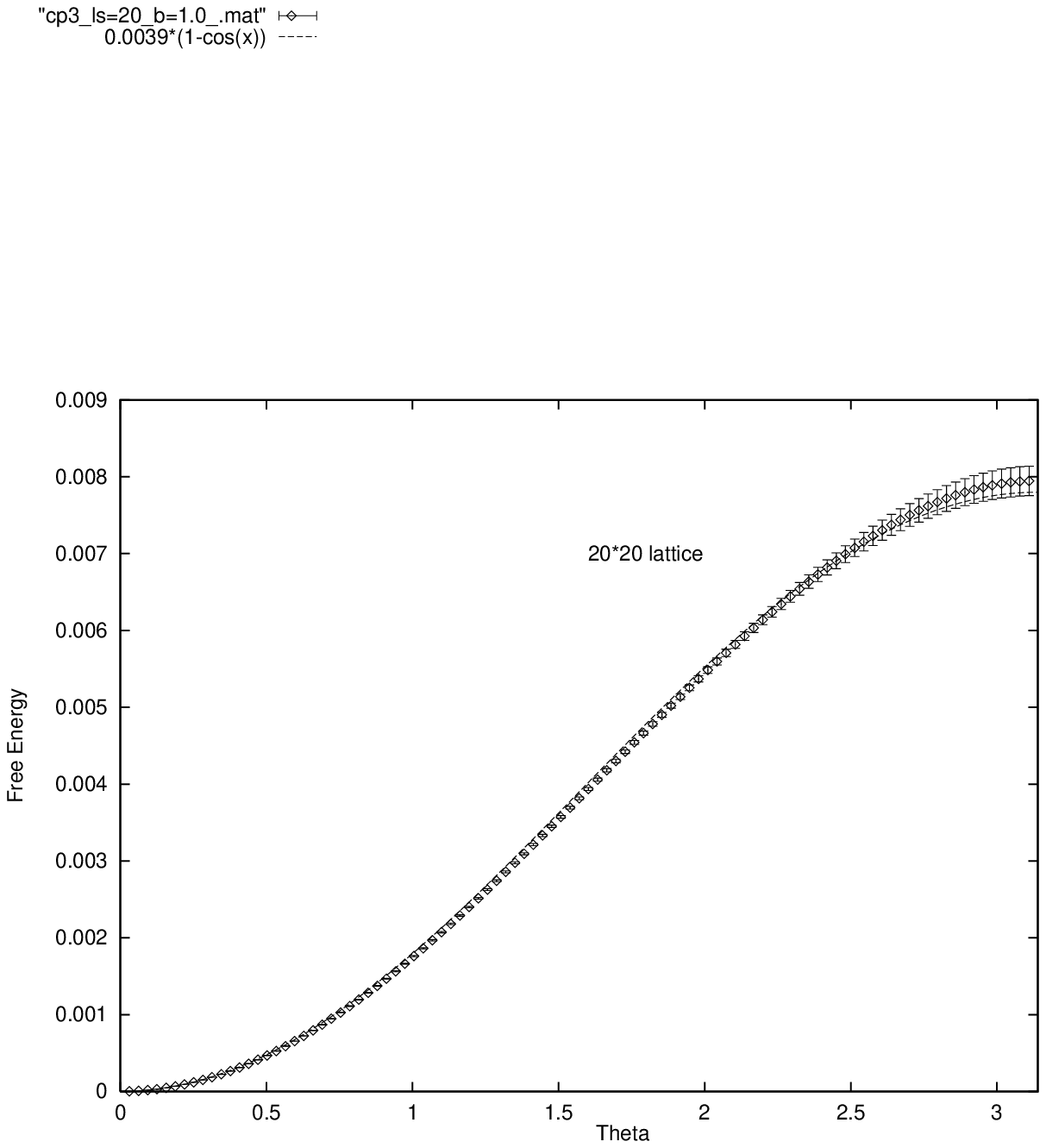}
\caption{$CP^3$ Free Energy Versus $\theta$ at $\beta=1.0$ 
on a $20\times 20$ Lattice and Its Cosine Fit.}
\end{figure}

\bigskip 
{\bf\large \noindent V.\ Remarks}  
\setcounter{section}{5}   
\setcounter{equation}{0}   

The cosine behavior of the free energy 
does not agree with the large $N$ limit result, 
which gives $f_{ {\rm large} \ N} = c/N \theta^2$, 
for some $N$-independent constant $c$.  
The discrepancy might be due to the smallness of $N$, 
which is $4$ in our case. 
In other words,  
higher-order $1/N$ corrections 
in the large $N$ expansion are important.  
We believe that this explanation is likely: 
In the leading zeroeth-order large $N$ expansion, 
no $\theta$ dependence arises 
in the free energy.  
The first $1/N$ correction provides 
the quadratic $\theta$ dependence. 
Hence, 
the first-order result differs significantly 
from the zeroeth-order result 
for large $\theta$ 
and 
especially  near $\theta = \pi$.  
It is thus quite possible that 
higher-order $1/N$ corrections 
contribute to the large $\theta$ region 
and cause the free energy to ``curve over'' 
like a cosine function.  
This would not be the first time that 
a large $N$ result is anomalous 
in the $CP^{N-1}$ models.\ct{hhr80,rs81,samuel83}  
In the zeroeth-order approximation, 
the $CP^{N-1}$ models do not confine.  
However, the first $1/N$ correction does 
lead to confinement.\ct{dlv78,witten79b} 
Nonetheless for finite $N$, 
the $CP^{N-1}$ models have a property called 
superconfinement \ct{samuel83} 
in which charges are so strongly bound 
that they cannot be separated by any non-zero distance.%
{\footnote{ 
An interesting side remark is that  
ref.\ \ct{ps96a} 
has shown that superconfinement 
is lost in the presence of a $\theta$ term.}}  
Superconfinement corresponds to an infinite string tension.  
It has been conjectured 
that summing the $1/N$ series  
would strengthen the confinement 
and lead to superconfinement.\ct{samuel83}  
Probably, quantities in the $CP^{N-1}$ models, 
which differ greatly 
in going from the zeroeth order to the first order, 
such as the free energy for $\theta$ away from $\theta = 0$
and the string tension, 
are not reliably computable in the large $N$ expansion.  

An interesting open question is whether there is 
a phase transition at $\theta=\pi$. 
At infinitely strong coupling, 
this is the case.\ct{seiberg84} 
In the strong coupling region, 
finite volume effects round off the cusp 
in the free energy 
when sufficiently small volumes are used 
to give reliable results near $\theta = \pi$. 
Hence definitive conclusions cannot be drawn. 
In the weak coupling region, 
the cosine function fits are suggestive 
that the phase transition is absent. 

If $f( \theta )  = c^{\prime} \theta^2$, 
then, as the volume $V$ goes to infinity, 
\be 
  { {P(Q)} \over {P(0)}} = 
   \exp \left( { - {{Q^2} \over {4 c^{\prime} V}} } \right) 
\quad .  
\label{eq5p1} 
\ee 
This follows from inverting 
Eq.\ \bref{eq2p4}:   
\be 
  { {P(Q)} \over {P(0)}} = 
 { { \int_{-\pi}^{\pi} {{d\theta} \over {2\pi}} 
        \exp ({ {i \theta Q} }) 
         \exp {( - V f( \theta ) )}  } 
       \over 
    { \int_{-\pi}^{\pi} {{d\theta} \over {2\pi}}  
        \exp {( - V f( \theta ) )} } 
  }
\quad .  
\label{eq5p2} 
\ee 
As $V \to \infty$, 
a saddle point expansion 
of Eq.\ \bref{eq5p2} 
becomes quite accurate. 
If the minimum of $f( \theta )$ is at $\theta = 0$, 
one finds 
\be 
  { {P(Q)} \over {P(0)}} = 
      \exp \left( { - a_2 {{Q^2} \over {2!}} - 
                      a_4 {{Q^4} \over {4!}} - \dots } \right)
\quad ,   
\label{eq5p3} 
\ee 
where 
$$
   a_2 = { {1} \over {V f^{(2)}(0)} } - 
         { {f^{(4)}(0)} \over {2 V^2 [f^{(2)}(0)]^3} } + 
         { {[f^{(4)}(0)]^2} \over {4 V^3 [f^{(2)}(0)]^5} } - 
         { {f^{(6)}(0)} \over { V^3 [f^{(2)}(0)]^4} } + 
           O({{1} \over {V^4}})
\quad ,   
$$ 
\be 
   a_4 = { {f^{(4)}(0)} \over {2 V^2 [f^{(2)}(0)]^4} } + 
            O( {{1} \over {V^4}}) 
\quad ,   
\label{eq5p4} 
\ee 
where $f^{(n)}(0)$ is the $n$th derivative of $f( \theta )$ 
at the origin.  
Equation \bref{eq5p3} 
reduces to Eq.\ \bref{eq5p1} 
if $f^{(n)}(0) = 0$ for $n>2$.  

There has been some previous discussion as to whether 
the $Q$ dependence in $P(Q)$ 
is gaussian.\ct{hity95a,hity95b}  
On one hand, 
Eq.\ \bref{eq5p3} 
shows that this is the case as $V \to \infty$.  
On the other hand, 
if the $Q$ dependence in $P(Q)$ is exactly gaussian 
then $f( \theta )  = c^{\prime} \theta^2$ 
for $\theta$ in the fundamental $\theta$ region, 
as $V \to \infty$.  
Since non-quadratic behavior for $f( \theta )$ 
is usually observed, 
the $Q^4$ and higher powers of $Q$ 
in Eq.\ \bref{eq5p3} 
are important for determining $f( \theta )$.  
Since the coefficient $a_4$ of $Q^4$ 
falls off as $1/V^3$, 
large volume systems make it difficult 
for Monte Carlo simulations to correctly determine 
the $\theta$ dependence of $f( \theta )$ 
for large $\theta$.  
One again arrives at the conclusion of guideline (2) 
of Section II:
Small volume results are to be trusted over large volume 
results for the computation of the free energy.  
For $\beta > 0.9$, 
the ``cosine'' behavior of $f( \theta)$ 
observed in the Monte Carlo investigations 
of the $CP^3$ model 
imply a non-gaussian behavior of $P(Q)$.  
In fact, 
in our simulations of both the $U(1)$ gauge model 
and of the $CP^3$ model, 
small derivations from gaussian behavior of $P(Q)$ 
were observed for all $\beta$ values.

\bigskip 
{\bf\large \noindent VI.\ Discussion}  
\setcounter{section}{6}   
\setcounter{equation}{0}   

By comparing Monte Carlo simulations 
of the $d=2$ $U(1)$ gauge theory, 
we have verified the guidelines 
given in Section II: 
For sufficiently large volumes, 
a barrier $\theta_b$ arises, 
beyond which numerical results 
for the free energy are unreliable. 
The limiting $\theta_b$
can be computed theoretically 
using Eq.\ \bref{estthetab} 
with $| \delta P(0) |$ estimated 
to be the statistical error in $P_{MC}(0)$.   
When the size of the system is reduced, 
reliable results are obtained 
throughout the entire fundamental region for $\theta$.  
Hence, when small-volume results differ 
from large-volume results, 
one should trust the smaller volume results.  
In some simulations, 
a flat free-energy behavior 
is observed for $\theta > \theta_b$.  
Comparison to exact analytic results 
demonstrates that the flat behavior 
is incorrect -- the true free energy 
continues to rise for $\theta > \theta_b$.  
The flattening effect can be attributed 
to the error in $P_{MC}(0)$ dominating 
over the errors in other $P_{MC}(Q)$.  
The domination of the error in $P_{MC}(0)$ 
is enhanced by Monte Carlo techniques 
such as the binning method 
and the use of a trial probability distribution.  

The above conclusions also hold 
for the ``auxiliary $U(1)$ field'' formulation 
of the $CP^3$ model:   
When the volume is large, 
anomalous flat behavior 
of the free energy 
is sometimes seen.  
At $\beta = 0.2$  
for the $6 \times 6$ lattice run, 
the flat-energy behavior 
is definitely incorrect 
since a comparison can be made 
with a reliable analytic strong-coupling expansion.   
When a smaller lattice size is used, 
results for the free energy are more accurate 
and flat behavior is absent.  
When the inverse coupling $\beta$ is 
greater than $0.6$, 
the strong-coupling calculation  
of the free energy has a peak, 
but the Monte Carlo data does not.   
See Figures 3 and 4.  
At these intermediate values of the coupling, 
we believe that the peak is an artifact 
of truncating the series to order ten --
higher-order contributions are probably important. 

There is a simple physical picture 
of why a limiting $\theta_b$ arises. 
Current methods for simulating a system with $\theta \ne 0$ 
are done using the $\theta = 0$ system.  
There should be a ``barrier'' separating the two systems.  
The barrier grows exponentially with the volume $V$.  
When $V$ is small or when $\theta$ is small, 
the barrier does not prevent the $\theta = 0$ system 
from sensing the physics of the $\theta \ne 0$ system. 
However, as $V$ gets large, 
the barrier becomes more impenetrable, 
and for sufficiently large $\theta$, 
the Monte Carlo simulations do not explore the phase space 
of the $\theta$-system sufficiently well to give reliable results.  

If one applies the guidelines in Section II 
to the work of ref.\ \ct{schierholz94}, 
one would conclude the following. 
The simulations of the adjoint form 
of the lattice $CP^3$ model in refs.\ \ct{schierholz94}  
found the absence of a flattening behavior 
in the free energy for sufficiently small volumes.  
This is typified in Figure 3 of the first of refs.\ \ct{schierholz94}.  
Guidelines (2) and (6) say that smaller-volume results 
are to be trusted over larger-volume results.  
Hence, 
one would conclude that the flattening behavior 
is anomalous.  
If this is true, 
it is a result of the probability method 
which tends to emphasize the error in $P_{MC}(0)$.  
If the flattening behavior is anomalous, 
$\theta_c$ of ref.\ \ct{schierholz94} 
should be identified 
with the barrier theta $\theta_b$.  
One test of this idea is as follows.  
Assuming that the statistical errors are 
approximately the same for all the runs 
in ref.\ \ct{schierholz94} 
and that Eq.\ \bref{eq1} 
holds, 
it follows 
from Eq.\ \bref{estthetab} 
that $V \theta_c^2$ should be 
approximately constant 
for a fixed value of $\beta$.
At $\beta = 2.5$, 
for the 
$28 \times 28$, 
$32 \times 32$,  
$38 \times 38$, and
$48 \times 48$ lattices,  
$V^{1/2} \theta_c /\pi$ 
is respectively 
$\sim 20$, $\sim 21$, $\sim 21$ and $\sim 22$.  
At $\beta = 2.7$, 
for the 
$56 \times 56$, 
$64 \times 64$, and 
$72 \times 72$ lattices,  
$V^{1/2} \theta_c /\pi$ 
is respectively 
$\sim 35$, $\sim 33$ and $\sim 32$.  
Hence, 
$V \theta_c^2$ is 
approximately constant. 
If the dominance of the error in $P_{MC}(0)$ 
is the source 
of the free-energy flattening behavior 
of refs.\ \ct{schierholz94,schierholz95}, 
one puzzling question arises: 
why did all the runs  
display flattening behavior -- 
one would have expected 
that, in some runs, 
the errors in the free energy 
to become large at $\theta_c$ 
and/or for the partition function 
to become negative.  
It is possible that some other subtle systematic effect 
is playing a role.  
The discussion here 
supports explanation (ii) 
of the Introduction, 
but one cannot definitively rule out 
explanation (i), namely,
that a massless or light mode arises 
for $\theta$ sufficiently large.

\medskip 
{\bf\large\noindent Acknowledgments}  

We thank Gerrit Schierholz and Peter Weisz for useful discussions.  
We acknowledge the Max-Planck Institute 
in Munich, Germany for its considerable support -- 
much of the computer work for this project 
was carried out on work stations at the Max-Planck Institute.   
We also thank the ITP of the University of Hanover
for use of its computers.  
This work was supported in part 
by the PSC Board of Higher Education at CUNY,   
by the National Science Foundation under the grant  
(PHY-9420615), 
and by two Humboldt Foundation grants, 
one of which is a Lynen-Fellowship.

\pagebreak

\def\NPB#1#2#3{ {Nucl.{\,}Phys.{\,}}{\bf B{#1}} ({#3}) {#2}} 
\def\PLB#1#2#3{ {Phys.{\,}Lett.{\,}}{\bf {#1}B} ({#3}) {#2}} 
\def\PRL#1#2#3{ {Phys.{\,}Rev.{\,}Lett.{\,}}{\bf  {#1}} ({#3}) {#2}} 
\def\PRD#1#2#3{ {Phys.{\,}Rev.{\,}}{\bf D{#1}} ({#3}) {#2}} 
\def\PR#1#2#3{ {Phys.{\,}Rep.{\,}}{\bf {#1}} ({#3}) {#2}} 
\def\OPR#1#2#3{ {Phys.{\,}Rev.{\,}}{\bf {#1}} ({#3}) {#2}} 
\def\NC#1#2#3{ {Nuovo Cimento{\,}}{\bf {#1}} ({#3}) {#2}}

\vfill\eject 
\end{document}